\documentclass[reprint,aps,prl,twocolumn,superscriptaddress,preprintnumbers,letterpaper,natbib,floatfix,nofootinbib]{revtex4-1}
\usepackage{amssymb}
\usepackage{amsmath}
\usepackage{epsfig}
\usepackage{comment}
\usepackage{color}
\usepackage{hyperref}
\usepackage{cleveref}
\usepackage{multirow}
\usepackage{capt-of}
\usepackage{siunitx}
\usepackage{graphicx}
\usepackage{slashed}
\usepackage{tabularx}
\usepackage{enumitem}
\usepackage{multirow}
\usepackage{booktabs}
\usepackage{isotope}
\usepackage{xcolor}
\usepackage{bm}
\usepackage{array,mathtools}

\newcommand{\Lag}{\mathcal{L}}

\newcommand{\prn}[1]{ \left(  #1 \right) }
\newcommand{\avg}[1]{\left< #1 \right>}

\newcommand{\sigmax}{\sigma^{\text{max}}_{\chi n}}
\newcommand{\TPT}{T_{\text{PT}}}

\newcommand{\Neff}{\text{N}_\text{eff}}
\newcommand{\al}[1]{\begin{align} #1 \end{align} }
\newcommand{\mysec}[1]{\paragraph*{#1.}\!\!\!\!\!---}
\newcommand{\Br}{\text{Br}}
\newcommand{\ctheta}[1]{\cos^{#1} \! \theta}
\newcommand{\bea}{\begin{eqnarray}}
\newcommand{\eea}{\end{eqnarray}}

\begin{document}
\title{Maximizing Direct Detection with Highly Interactive Particle Relic Dark Matter}
\author{Gilly Elor}
\affiliation{\sl PRISMA$^+$ Cluster of Excellence \& Mainz Institute for Theoretical Physics\\
Johannes Gutenberg University, 55099 Mainz, Germany\\[2mm]}
\author{Robert McGehee}
\author{Aaron Pierce}
\affiliation{Leinweber Center for Theoretical Physics, Department of Physics,\\ University of Michigan, Ann Arbor, MI 48109, USA}

\begin{abstract}
We estimate the maximum direct detection cross section for sub-GeV dark matter (DM) scattering off nucleons. For DM masses in the range $10 \text{ keV }- 100 \text{ MeV}$, cross sections greater than $10^{-36}$- $10^{-30} \,\text{cm}^2$ seem implausible. We present a DM candidate which realizes this maximum cross section: HighlY interactive ParticlE Relics (HYPERs). After HYPERs freeze-in, a dark sector phase transition decreases the mediator's mass. This increases the HYPER's direct detection cross section without impacting its abundance or measurements of Big Bang Nucleosynthesis and the Cosmic Microwave Background. 
\end{abstract}

\maketitle
\preprint{LCTP-21-26}
\preprint{MITP-21-044}

In the face of null results in the direct search for Weakly Interacting Massive Particle (WIMP) dark matter (DM), the motivation to explore alternative DM candidates has steadily grown. One possibility is to explore DM models with even smaller interactions, which requires experiments with larger exposures and exquisite control of backgrounds. Another possibility is to instead consider larger interactions for DM models 
too light to be directly detected at current experiments. But what is the maximum cross section for sub-GeV DM $\chi$ scattering off nucleons, $\sigmax$? This is the first question we address.

The second is inspired by the proliferation of proposals for direct detection experiments sensitive to sub-GeV DM coupled to nucleons (see \emph{e.g.}~\cite{Knapen:2016cue,Budnik:2017sbu,Knapen:2017ekk,Griffin:2018bjn,Kurinsky:2019pgb,Essig:2019kfe,Trickle:2019nya,Griffin:2019mvc,Campbell-Deem:2019hdx,Griffin:2020lgd,Coskuner:2021qxo}). While $\sigmax$ will provide an important guidepost for these proposals, it is desirable to have examples of DM models which have a consistent cosmology.\footnote{In contrast, experiments sensitive to electron couplings, including many of the above cited proposals, often probe well motivated freeze-in and freeze-out benchmarks.} Is there a sub-GeV DM candidate which may be detected at these future DM-nucleon scattering experiments? Could it have a cross section as large as $\sigmax$ while still accounting for its relic abundance and cosmological history? 

A large $\sigma_{\chi n}$ would arise if $\chi$ were to interact with a light mediator $\phi$ with sizable couplings to both the DM and nuclei \cite{Knapen:2017xzo}. However, a thermal history for such a scenario could suffer from two  challenges. First, a large annihilation rate $\bar{\chi} \chi \to \phi \phi$ could deplete the DM relic abundance in the early Universe. These fast annihilations could also be constrained by present-day indirect detection bounds \cite{Laha:2020ivk,Boddy:2015efa}. Second, thermalization of the light $\phi$ could increase $\Neff$ \cite{Krnjaic:2019dzc,Sabti:2019mhn}, in tension with measurements. While the $\phi$ must be light today to ensure the large direct detection cross section, if it had a heavier mass at earlier times, it might mitigate these challenges.

With this motivation, we introduce a new DM candidate: \emph{HighlY interactive ParticlE Relics} (HYPERs). HYPERs are designed to evade these dangerous cosmological bounds, and are thus a candidate for realizing $\sigmax$. HYPER models refer to DM scenarios in which the mediator mass drops after the DM relic abundance is determined, thus boosting the present-day interactions between the DM and the Standard Model (SM). After the $\chi$'s relic abundance is set, a phase transition in the dark sector causes the mediator, which connects the DM to the visible sector, to decrease in mass to its present-day value\footnote{Other models have considered a dark sector phase transition changing particle masses (see \emph{e.g.} \cite{Cohen:2008nb,Baker:2016xzo,Croon:2020ntf,Hashino:2021dvx}, see also \cite{Dimopoulos:1990ai}) or interactions much stronger today than when the relic abundance was set (\emph{e.g.}~\cite{Boddy:2012xs}).} $m_\phi^i \to m_\phi$. HYPER direct detection cross sections are thus enhanced by a factor $(m_\phi^i/m_\phi)^4$. 

In this Letter, we first estimate $\sigmax$ for sub-GeV DM by considering only present-day experimental and astrophysical constraints, without making reference to cosmology. We develop a simple, hadrophilic HYPER model which can realize $\sigmax$ for some DM masses while avoiding cosmological bounds. We then detail the HYPER (parameter) space, highlighting  regions in which our hadrophilic HYPER model can reach $\sigmax$. This exercise also highlights the difficulties in constructing models with large direct detection cross sections. 

\mysec{Estimating $\sigmax$}To estimate $\sigmax$, we assume a scalar mediator $\phi$ is coupled to DM and nucleons, $n$:
\al{
\label{eq:Lagrangian}
\Lag \,\,\supset\,\, -m_\chi \bar{\chi} \chi -y_n \phi \bar{n} n - y_\chi \phi \bar{\chi} \chi \,.
}
This gives a maximum direct detection cross section 
\begin{align}
    \sigmax \equiv \frac{\prn{y_n^{\text{max}} y_\chi^\text{max}}^2}{ \pi} \frac{\mu_{\chi n}^2}{\left[(m_\phi^\text{min})^2 + v_{\chi}^2 m_\chi^2\right]^2}\,.
\end{align}

\begin{figure}[t!]
\centering
\includegraphics[width=0.98 \columnwidth]{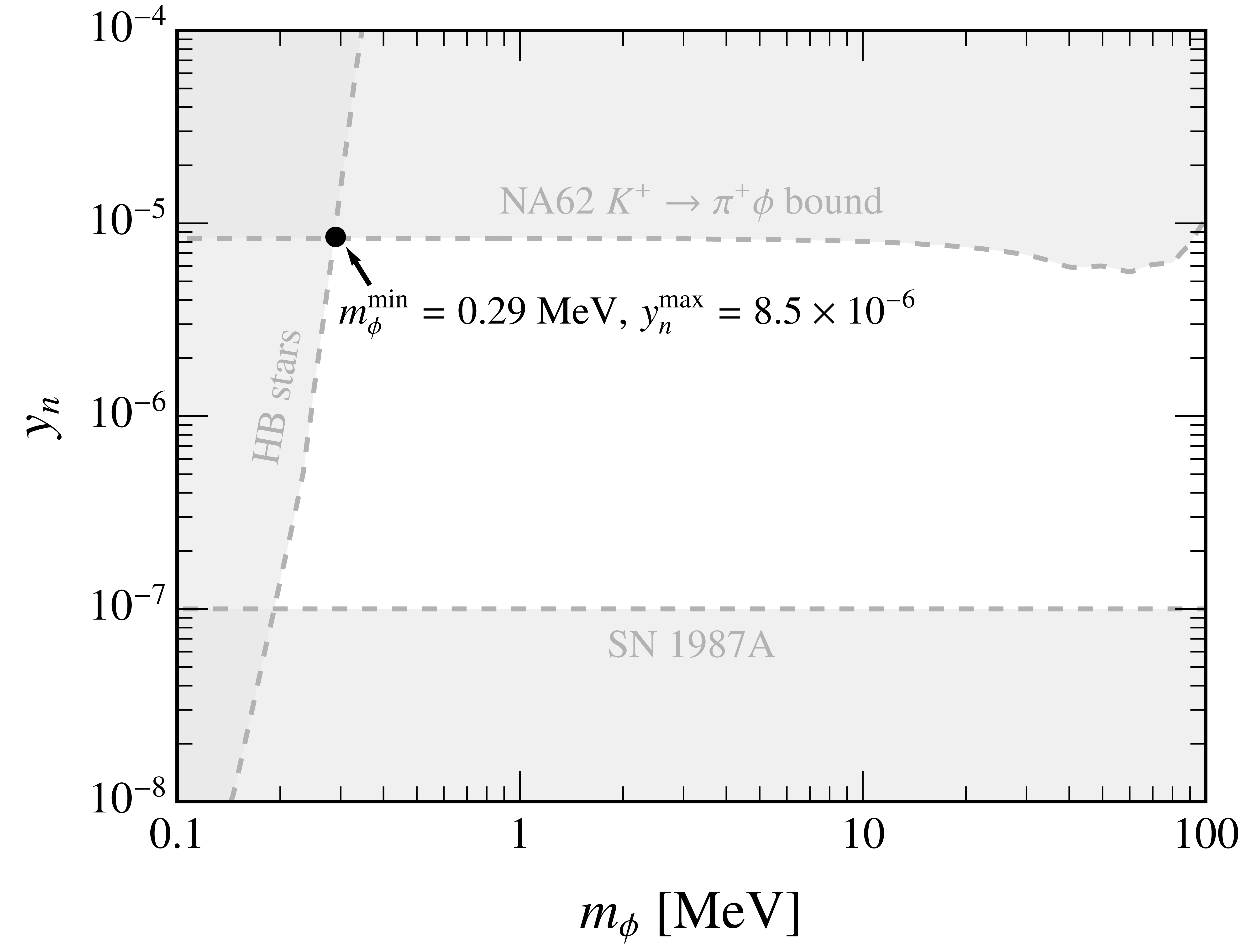}
\caption{Constraints in the mediator mass -- nucleon coupling plane from cooling of HB stars~\cite{Hardy:2016kme} and SN 1987A~\cite{Knapen:2017xzo}, as well as rare Kaon decays~\cite{NA62:2021zjw} (grey shading). Also shown are values for $(m_\phi^\text{min},y_n^\text{max})$. }
\label{fig:ynmphibnd}
\end{figure}

The first step in estimating $\sigmax$ in this model is to obtain the extremal values of $m_\phi$, $y_n$, and $y_\chi$ consistent with \emph{present-day} bounds. There are a range of mediator-nucleon couplings $y_{n}$ for which $m_{\phi} \gtrsim 0.3 $ MeV prevents disturbing the dynamics of Horizontal Branch (HB) stars while avoiding constraints from supernova (SN) cooling. The bounds on $y_n$ depend on its origin. One possibility is that it arises from a coupling to gluons, $\phi G^{\mu \nu} G_{\mu \nu}$. This coupling can be generated upon integrating out heavy colored degrees of freedom, as could happen if  $\phi$ couples to top quarks or to new heavy, vectorlike quarks, $\psi$ \cite{Knapen:2017xzo}. Constraints that arise from rare decays of mesons are weaker in the latter scenario, so we assume this UV completion to maximize $\sigma_{\chi n}$. While the nucleon coupling could arise from couplings to light quarks~\cite{Batell:2018fqo,Ema:2020ulo}, such set-ups are likely to be even more susceptible to bounds from meson decays unless the mediator is heavier than $\mathcal{O}(1\text{ GeV})$, which would substantially suppress direct detection cross sections. So, we specialize to the case where the nucleon coupling arises from a gluon coupling that comes from integrating out vectorlike quarks. 

\begin{figure}[t!]
\centering
\includegraphics[width=0.98 \columnwidth]{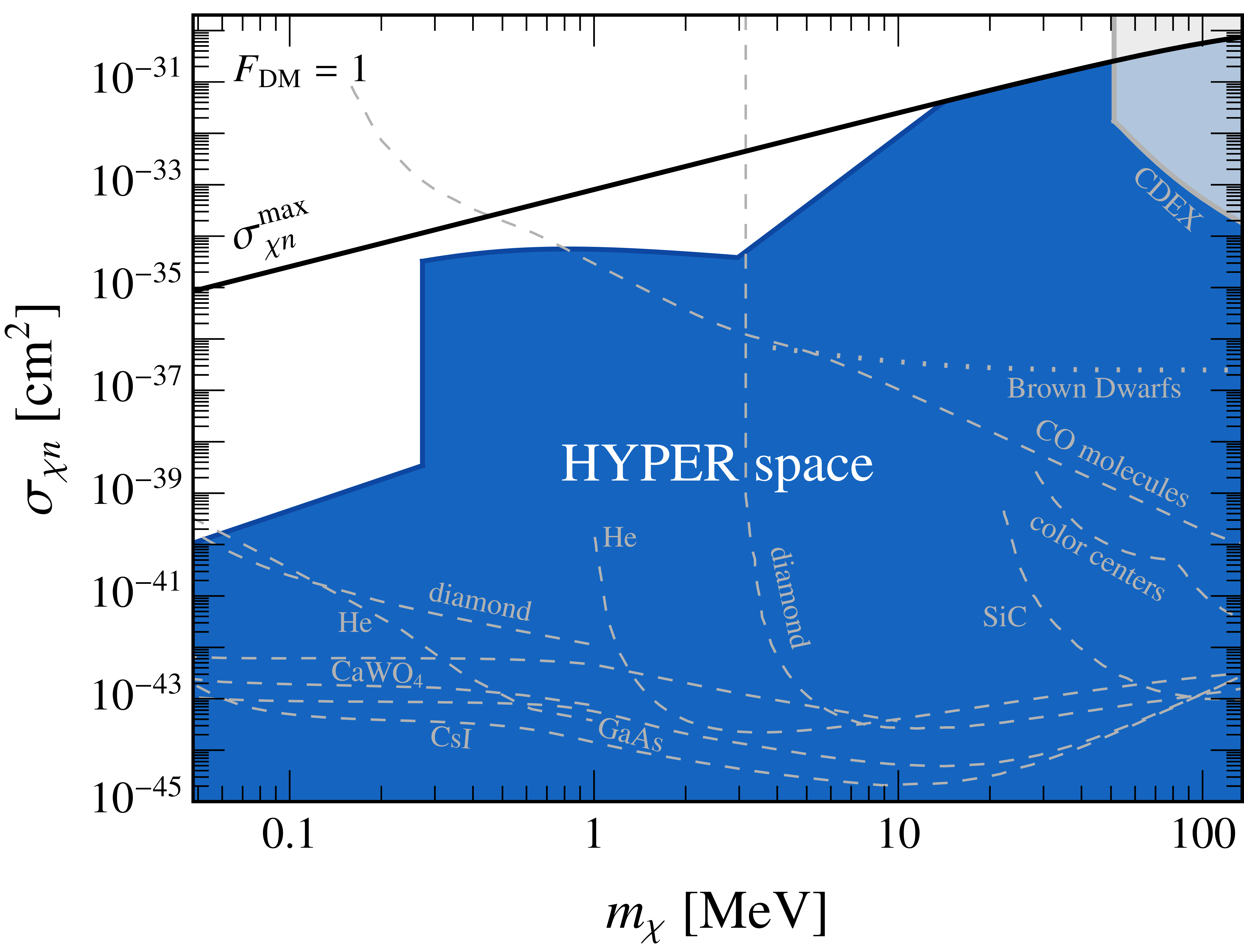}
\caption{$\sigmax$ for the $(m_\phi^\text{min},y_n^\text{max})$ values from Fig.~\ref{fig:ynmphibnd}, as well as hadrophilic HYPER space for the $\TPT = 1 \text{ MeV}$ benchmark. The current constraint from CDEX~\cite{CDEX:2019hzn} is shaded gray, while future projected sensitivities are shown with dashed gray lines \cite{Knapen:2016cue,Budnik:2017sbu,Knapen:2017ekk,Griffin:2018bjn,Kurinsky:2019pgb,Essig:2019kfe,Trickle:2019nya,Griffin:2019mvc,Campbell-Deem:2019hdx,Griffin:2020lgd,Coskuner:2021qxo,Leane:2020wob}.} 
\label{fig:sigmax}
\end{figure}

The relevant bounds on $m_\phi$ vs. $y_n$ are shown in gray in Fig.~\ref{fig:ynmphibnd}. They include cooling bounds from supernova (SN) 1987A~\cite{Knapen:2017xzo} and HB stars~\cite{Hardy:2016kme}. The vectorlike colored fermions induce a $\phi \bar{t} t$ coupling at two loops, which for fixed $y_{n}$ depends logarithmically on the mass of the vectorlike fermions. This induces $K^+ \to \pi^+ \phi$ decay at loop-level via CKM mixing \cite{Knapen:2017xzo}. To minimize the size of the $\phi \bar{t} t$ coupling while avoiding LHC bounds, we take the mass $m_{\psi}$ of this vectorlike fermion to be 1.5 TeV.  The induced Kaon decay is bounded by limits on $\Br \prn{K^+ \to \pi^+ X}$, where $X$ is an invisible spin-0 particle \cite{NA62:2021zjw}. The result is the ``NA62 $K^+ \to \pi^+ \phi$ bound'' in Fig.~\ref{fig:ynmphibnd}. Future data from NA62 will strengthen the bound on $y_n$ and hence decrease $\sigmax$. 

Next, we maximize $y_\chi$. The strongest bound comes from DM elastic scattering. We use the Born approximation to the transfer cross section given in~\cite{Knapen:2017xzo} and saturate the bound on the self interaction cross section $\sigma_{\chi \chi}/m_\chi \lesssim 1 \text{ cm}^2/\text{g}$ at $v_{\text{DM}} \sim 10^{-3}$ \cite{Kaplinghat:2015aga} to find $y_\chi^\text{max}$. 

We have verified that indirect detection bounds on $y_{\chi}$ are significantly weaker. There are two processes particularly worth checking.  First, $\phi$ can mediate DM  annihilations to photons. While non-perturbative quark and gluon loops prevent a precise calculation of the $\phi$ coupling to photons, a naive dimensional analysis estimate  yields 
\al{
\label{eq:LphiFF}
\Lag \, \supset \, \frac{\alpha y_n}{4 \pi m_n} \phi F_{\mu \nu} F^{\mu \nu}\,.
}
This permits DM annihilations to photons $\bar{\chi} \chi \to \gamma \gamma$ in the center of galaxies,  with a $p$-wave cross section given in the Appendix. When evaluated using the virial velocity of the Milky Way, this cross section is roughly 10 orders of magnitude smaller than the bound \cite{Laha:2020ivk}. Second, $\chi$ annihilation into a pair of on-shell $\phi$'s, which could subsequently decay to photons, could be subject to indirect detection constraints. While the precise bound depends on the details of the photon spectra as determined by the $m_{\chi}/m_{\phi}$ mass ratio, even in the most constraining case when the emitted photons are monochromatic, recasting bounds from \cite{Laha:2020ivk,Boddy:2015efa}, we find $y_\chi^{\rm max}$ is unaffected above $m_\chi = 10 \, \text{MeV}$. For smaller $m_\chi$, we find a slightly smaller $y_\chi^\text{max}$. However, these bounds are easily avoided if the $\phi$'s could decay to a light dark state. The branching ratio to the hidden state can be large even without introducing large couplings between $\phi$ and the light dark state, since the induced coupling in Eq.~\eqref{eq:LphiFF} is very small. Therefore, in our determination of $\sigmax$, we do not impose a constraint coming from present-day $\bar{\chi} \chi \to \phi \phi$ annihilations.

We show our $\sigmax$ estimate in Fig.~\ref{fig:sigmax} using the values from Fig.~\ref{fig:ynmphibnd}. For DM masses in the range of $10 \text{ keV }- 100 \text{ MeV}$, cross sections greater than $10^{-36}$- $10^{-30} \,\text{cm}^2$ seem implausible. We show the current constraint from China Dark Matter Experiment (CDEX) \cite{CDEX:2019hzn} (shaded gray), along with the projected sensitivities of future experiments (assuming one kg-year exposure) in dashed gray \cite{Knapen:2016cue,Budnik:2017sbu,Knapen:2017ekk,Griffin:2018bjn,Kurinsky:2019pgb,Essig:2019kfe,Trickle:2019nya,Griffin:2019mvc,Campbell-Deem:2019hdx,Griffin:2020lgd,Coskuner:2021qxo}. Brown Dwarfs may also probe large cross sections (dotted gray) \cite{Leane:2020wob}. It is noteworthy that all of these proposals are sensitive to the $\sigmax$ curve and therefore, capable of discovering DM. No bounds related to cosmic ray scattering \cite{Cappiello:2018hsu,Bondarenko:2019vrb,PROSPECT:2021awi} appear in Fig.~\ref{fig:sigmax}. They are orders of magnitude above the $\sigmax$ line and are unlikely to constrain point-like DM models.\footnote{Stronger bounds/projections in \cite{Bringmann:2018cvk,Cappiello:2019qsw} 
assume unphysical, constant (with respect to energy) cross sections.} 

We now comment on the robustness of $\sigmax$ against variations of our starting assumptions. If we permit sufficient fine tuning when UV completing the nucleon coupling in Eq.~\eqref{eq:Lagrangian}, the constraints in Fig.~\ref{fig:ynmphibnd} can be weakened. For instance, the NA62 bound on Kaon decays may be avoided by adding a term to the Lagrangian $\propto \phi QHu_R \to \phi \bar{t} t$ which cancels the loop induced contribution coming from the heavy, vectorlike colored fermions. 

One may also wonder if a larger $\sigmax$ could be achieved with a vector mediator. We find that current beam dump and collider bounds \cite{Alexander:2016aln,Parker:2018vye} on visibly decaying dark photons and $U(1)_{B-L}$ vectors result in smaller cross sections over the range of $m_\chi$ we consider. HB stars bound invisibly decaying dark photon masses similarly to $m_\phi^\text{min}$~\cite{Redondo:2013lna,Hardy:2016kme}, and constraints on DM self-scattering bound the dark gauge coupling, $g_D \lesssim y_\chi^\text{max}$. The kinetic mixing between the dark and SM photons 
is bounded to be roughly of order $y_n^{\rm max}$ \cite{Danilov:2018bks}, resulting in a $\sigmax$ similar to ours. All of these anomaly-free vectors couple to leptons as well as baryons. This results in bounds at light vector masses not present in the scalar mediator case. Gauging anomalous symmetries, e.g. baryon number, will involve challenging UV completions, and also result in ``anomalon'' bounds \cite{Dror:2017ehi}. Satisfying these bounds results in a cross section smaller than our $\sigmax$ by more than 6 orders of magnitude. 

Finally, one may consider composite instead of point-like DM~\cite{Digman:2019wdm}. If it is asymmetric with sufficiently large composite states, it may have an enhanced direct detection cross section \cite{Coskuner:2018are}. 

\mysec{A hadrophilic HYPER model}With an estimate of $\sigmax$ in hand, the next question is: are there DM models with such a large cross section that explain the relic abundance?
Because of crossing symmetry, such highly interactive DM would be expected to over-annihilate in the early Universe and consequently have too small a relic abundance. HYPERs avoid this problem by having a sufficiently late dark sector phase transition and their relic abundance set by UV freeze-in\footnote{It may also be possible to probe freeze-in DM at the LHC or in large direct detection experiments (see \emph{e.g.}~\cite{Co:2015pka,Hessler:2016kwm,Belanger:2018sti,Barman:2021lot}).} \cite{Hall:2009bx,Elahi:2014fsa}: $\chi$ and $\phi$ never come into thermal equilibrium with the SM. This also prevents an increase in $N_{\rm eff}$ \cite{Krnjaic:2019dzc,Sabti:2019mhn}. 

The range of HYPER masses we consider is 
\al{\label{eq:mchirange}
\mathcal{O}(10 \text{ keV})\lesssim m_\chi < m_{\pi^0}\,.
}
In this range---also of interest for the experimental proposals in Fig.~\ref{fig:sigmax}---it is easier to build models that approach $\sigmax$. The upper bound kinematically forbids $\bar{\chi} \chi \to \text{ hadrons}$ when $T \lesssim m_{\pi^0}$ which could reduce the DM abundance once a large coupling to nucleons is assumed. The lower bound ensures consistency with bounds from Lyman-$\alpha$ measurements  \cite{Ballesteros:2020adh}, though more stringent bounds exist for IR freeze-in \cite{DEramo:2020gpr}.

We now outline the HYPER thermal history. The reheat temperature, $T_R$, is much greater than the temperature of the dark sector phase transition but below $m_\phi^i$. As discussed above, a UV completion of Eq.~\eqref{eq:Lagrangian} using vectorlike quarks permits a larger $y_n$, and so we assume this scenario for our hadrophilic HYPER model. Integrating out both the heavy vectorlike quark $\psi$ and the initially heavy mediator $\phi$, leads to an effective operator: $\frac{\alpha_s y_\chi y_n}{2.8\,m_n (m_\phi^i)^2}\bar{\chi}\chi G^{a,\mu \nu} G^a_{\mu \nu}$. HYPERs thus freeze-in through this dimension-7 coupling to gluons. This effective operator description is consistent if $T_R \lesssim \text{Min} \left[m_\phi^i/20 \,, m_\psi/20 \right]$, where the factor of 20 Boltzmann suppresses $\phi$s or $\psi$s, so that production of DM via these states may be neglected.  HYPERs will then be mainly produced at the temperature $T_{R}$, i.e., HYPERs UV freeze-in with yield \cite{Elahi:2014fsa} 
\bea
Y_\text{DM} \,\simeq\, 5.3 \left( \frac{ y_n y_\chi \alpha_s }{m_n (m_\phi^i)^2} \right)^2 \frac{M_{\rm Pl} T_R^5}{g_{s,*} \sqrt{g_*}}\,. 
\eea
Both $T_R$ and $m_\phi^i$, which we adjust to obtain the correct relic abundance: $Y_\text{DM} m_\chi \simeq 4.4 \times 10^{-10} \, \text{GeV}$ \cite{Planck:2018vyg}, have no impact on the HYPER's final direct detection cross section. The $y_n^{\rm max}$ in Fig.~\ref{fig:ynmphibnd} corresponds to $m_\psi= 1.5 \, \text{TeV}$, which in turn requires $T_R \lesssim 75 \text{ GeV}$. We have checked that reheat temperatures from 18 GeV up to 75 GeV may result in the correct HYPER abundance over the range of masses in Eq.~\eqref{eq:mchirange} and for the couplings $y_n^\text{max}$ and $y_\chi^\text{max}$.

As the Universe cools, the dark sector undergoes a phase transition when the SM bath temperature reaches $\TPT$ which results in a significant drop in mass for the mediator $m_\phi^i \to m_\phi \ll m_\phi^i$. \footnote{If $\chi$ were a scalar, one might expect its mass to change as a result of this phase transition too. For simplicity, we have assumed $\chi$ is a fermion.} While this transition has the effect of increasing the direct detection cross section, we must ensure that this mass drop does not change the DM abundance or lead to new cosmological constraints on the mediator. We assume that any additional dark sector particles and couplings necessitated by the phase transition do not affect the DM energy density or abundance. While the requirement $\TPT \ll m_\pi$ avoids cosmological constraints involving hadrons, there are still processes which need to be sufficiently slow after the phase transition to avoid appreciably changing the DM abundance: $\gamma \gamma \to \phi$, $\bar{\chi} \chi \to \gamma \gamma$, $\gamma \gamma \to \bar{\chi} \chi$, and $\bar{\chi} \chi \to \phi \phi$.

Inverse decays of pairs of photons to $\phi$s are harmless as long as $2m_\chi > m_\phi$ since $\phi$s cannot decay to DM pairs and instead just decay harmlessly back to photons. Thus, $\gamma \gamma \to \phi$ only matters when $2m_\chi < m_\phi$ and we must increase $m_\phi$ to sufficiently prevent $\gamma \gamma \to \phi$, as we discuss in the Appendix.

The requirement that DM annihilations to photons after the phase transition do not deplete its abundance is
\al{
\sigma v_{\bar{\chi} \chi \to \gamma \gamma} n_\chi^2 \lesssim 3 H n_\chi\,,
}
where $\sigma v_{\bar{\chi} \chi \to \gamma \gamma}$ is given by Eq.~\eqref{eq:chichi2gamgam}. This is satisfied by many orders of magnitude for HYPERs with maximized couplings and minimum mediator mass.

The cross section for the reverse process
$\gamma \gamma \to \bar{\chi} \chi$ is
\al{
\sigma v_{\gamma \gamma \to \bar{\chi} \chi} = \frac{y_\chi^2}{2\pi} \prn{\frac{y_n \alpha}{4 \pi m_n}}^2 \frac{T \prn{T^2-m_\chi^2}^{3/2}}{\prn{T^2-m_\phi^2/4}^2}\,,
}
where we have assumed each photon has energy of the order $T$. The requirement that this process does not appreciably produce DM after the phase transition is
\al{
\sigma v_{\gamma \gamma \to \bar{\chi} \chi} n_\gamma^\text{eq} n_\gamma^\text{eq} \lesssim 3 H n_\chi \,,
\label{eq:gamgam2chichiCond}
}
where everything is evaluated at $T=\TPT$ since the condition is hardest to satisfy at greater temperatures. 

The final potentially troublesome process after the phase transition is $\bar{\chi} \chi \to \phi \phi$ (see Appendix for cross section).   At temperatures above $\TPT$, this is kinematically forbidden since $m_{\phi}^i$ is greater than both $T_R$ and $m_\chi$. But after the phase transition (at $\TPT$),  we must check that
\al{
\label{eq:chichi2phiphiCond}
\sigma v_{\chi \bar{\chi} \rightarrow \phi \phi} n_\chi < 3H\,.
}
We find that HYPERs which satisfy Eq.~\eqref{eq:chichi2phiphiCond} also avoid present-day indirect detection bounds coming from $\bar{\chi} \chi \to \phi \phi$, followed by $\phi$ decays to gamma rays.\footnote{As discussed in the appendix, this process is $p$-wave and does not suffer from CMB constraints \cite{Slatyer:2015kla,Liu:2016cnk}.}

We note that inverse decays $\bar{\chi} \chi \to \phi$ are innocuous--if active, the produced $\phi$s would promptly decay back to DM pairs and not change the HYPER abundance. For some $m_\chi$ and $\TPT$, satisfying Eqs.~(\ref{eq:gamgam2chichiCond}-\ref{eq:chichi2phiphiCond}) requires HYPERs to have $m_\phi > m_\phi^\text{min}$ and/or $y_\chi < y_\chi^\text{max}$, as we show next.

\mysec{Maximizing Direct Detection with HYPERs}Having estimated $\sigmax$ and introduced the hadrophilic HYPER model, we now illustrate how HYPERs can achieve $\sigmax$ as a proof of concept. Eqs.~(\ref{eq:gamgam2chichiCond}-\ref{eq:chichi2phiphiCond}) make this exercise non-trivial and help demonstrate the challenges of constructing a model with cross sections approaching $\sigmax$.

As the dark phase transition temperature increases, there are a greater number of processes that could potentially impact the DM abundance following the transition which must be suppressed. This suppression comes at the price of also suppressing $\sigma_{\chi n}$, opposite our goal of discovering how close HYPERs can get to $\sigmax$. Thus, we choose the benchmark $\TPT = 1 \, \text{MeV}$. At this temperature, it seems possible to evade the most stringent bounds coming from BBN on $\Neff$ or disturbing the deuteron abundance. We also require by fiat that the phase transition not impact the frozen-in relic abundance.

With these considerations in mind, we now present the $\TPT=1$ MeV benchmark HYPER space in Fig.~\ref{fig:sigmax}. Everywhere, we choose model parameters $(m_\phi, y_\chi)$ to maximize $\sigma_{\chi n}$ while evading the processes discussed above. This is done by setting $y_\chi < y_\chi^{\rm max}$ or $m_\phi > m_\phi^{\rm min}$ or both, when necessary. Everywhere, $y_n=y_n^\text{max}$. Once these constraints are satisfied, HYPERs may be chosen to have $y_\chi$ smaller than the value which maximizes $\sigma_{\chi n}$,  and there exists a continuous deformation from the 1 MeV benchmark HYPERs to ordinary models of UV freeze-in. The range of HYPER models for $\TPT = 1 \, \text{MeV}$ in $(m_\chi,\sigma_{\chi n})$ is shaded and extends all the way down to the UV freeze-in line (with no phase transition), which varies from $10^{-66}$-$10^{-62} \text{ cm}^{2}$ over the range of DM masses shown for $T_R=75 \text{ GeV}$.

The boundary of HYPER space is determined by choosing model parameters $(m_\phi, y_\chi)$ that maximize $\sigma_{\chi n}$ at each DM mass. For $m_\chi > 14 \text{ MeV}$, HYPERs succeed in saturating $\sigmax$ because $\bar{\chi} \chi \to \phi \phi$ is $p$-wave suppressed, and heavier $m_\chi$ have smaller velocities at $\TPT$. This allows us to set $m_\phi = m_\phi^\text{min}$ and $y_\chi = y_\chi^\text{max}$ while still satisfying the constraint in Eq.~(\ref{eq:chichi2phiphiCond}). For $3.0 \text{ MeV} < m_\chi < 14 \text{ MeV}$, this $p$-wave suppression is insufficient, and $\bar{\chi} \chi \to \phi \phi$ must be suppressed. The largest $\sigma_{\chi n}$ is achieved by setting $m_\phi = m_\phi^\text{min}$ and suppressing $y_\chi < y_\chi^\text{max}$ sufficiently to satisfy Eq.~\eqref{eq:chichi2phiphiCond}. For $270 \text{ keV} < m_\chi < 3.0 \text{ MeV}$, $\bar{\chi} \chi \to \phi \phi$ is still the most problematic process after $\TPT$. However, the largest $\sigma_{\chi n}$ is instead achieved by kinematically forbidding this process: we set $m_\phi=E_\chi(\TPT)$ (and above $m_\phi^\text{min}$). Here, $E_\chi(\TPT)$ is the energy of a HYPER at the phase transition, and we take into account the dilution of their kinetic energy relative to the SM bath temperature due to many degrees of freedom leaving the SM bath after $T_R$. 

At $m_\chi = 270 \text{ keV}$, there is a sharp drop in the largest $\sigma_{\chi n}$. At this $m_\chi$, setting $m_\phi=E_\chi(\TPT)$ results in $m_\phi \ge 2 m_\chi$ for all smaller $m_\chi$. Therefore, $\gamma \gamma \to \phi$ inverse decays pose a serious threat of increasing the DM relic abundance through subsequent $\phi \to \bar{\chi} \chi$ decays. The best way to prevent these inverse decays from producing an appreciable amount of DM while keeping a detectable $\sigma_{\chi n}$ is to increase $m_\phi$ to Boltzmann suppress this process. We find that the inverse decays produce less than $10\%$ of the DM abundance for $m_\phi = 21 \text{ MeV}$, and we then set $y_\chi=y_\chi^\text{max}$; we have checked Eq.~\eqref{eq:gamgam2chichiCond} is satisfied. The lightest $m_\chi$ we consider is $m_\chi = \omega_p(\TPT)/2=48 \text{ keV}$, where $\omega_p(\TPT)$ is the plasma frequency at the phase transition. It is possible that longitudinal plasmons mixing with $\phi$ may allow for $\gamma^\ast \to \bar{\chi} \chi$ decays \cite{Dvorkin:2019zdi}. Careful exploration of this process is postponed to future work.

In summary, HYPERs can have direct detection cross sections as large as $\sigmax$ over roughly an order of magnitude in mass, for $14 \text{ MeV} < m_\chi < m_{\pi^0}$, when $\TPT = 1 \text{ MeV}$.   Our estimate of $\sigmax$ did not consider DM's relic abundance or cosmological history.  It is non-trivial that there exists a cosmological story such as HYPERs which can not only achieve $\sigmax$ while evading the usual early-Universe constraints, but can do so and still explain the DM relic abundance. 

The cost of a large direct detection signal for HYPERs seems to be one of fine tuning in the dark sector scalar potential. For the models with the largest $\sigma_{\chi n}$, the potential must cause $m_\phi^i \sim \mathcal{O}\prn{100 \, T_R} \to m_\phi \sim \mathcal{O}\prn{\text{MeV}}$  at a late-time phase transition close to $\TPT \sim \mathcal{O}\prn{\text{MeV}}$. To do this for $T_R \sim 75\, \text{GeV}$ without contributing a sizeable vacuum energy or entropy dump to the SM at $\TPT$ requires a quite flat direction in the scalar potential.  

Additionally, the phase transition requires a large vacuum expectation value (VEV) at high temperatures to transition to a much \emph{smaller} VEV at lower temperatures. While transitioning from a large value to a small value could occur in simple potentials with two-step phase transitions \cite{Kurup:2017dzf,Chiang:2017nmu}, the presence of such disparate energy scales is a significant challenge for model building. 

Note that a dark sector phase transition that occurs after or during BBN is not a priori excluded, and if viable, a $\TPT < m_\chi$ would kinematically forbid most of the dangerous processes that can occur after $\TPT$.  This would allow the upper edge of HYPER space to move closer to $\sigmax$. In principle, if the change in the potential is sufficiently small, one could evade BBN and CMB constraints \cite{Bai:2021ibt,Breitbach:2018ddu}, and we leave this and other phase transition investigations to future work \cite{future}. 

\mysec{Discussion}In this Letter, we have addressed questions relevant to the search for sub-GeV DM. 1) We have estimated $\sigmax$ for DM coupled to nucleons. In particular, we find cross sections greater than about $10^{-36}-10^{-30}\, \text{cm}^2$ for DM masses $10 \text{ keV } < m_{\chi} < 100 \text{ MeV}$ are implausible. 2) We have introduced a new type of DM with cross sections as large as $\sigmax$, \emph{HYPERs}. HYPERs populate a parameter space which is imminently testable by future direct detection efforts but has few DM benchmarks. 

\newpage
It would be interesting to see what kinds of hadrophilic DM models other than HYPERs could (nearly) saturate $\sigmax$.\footnote{Point-like asymmetric DM may still have a cross section as large as $\sigma^{\rm max}_{\chi n}$ for $m_\chi \gtrsim \mathcal{O}(10 \text{ MeV})$. However, a mediator with mass $m_{\phi}^{\rm min}$ would be in slight tension with BBN bounds on $\Neff$ without further model building \cite{Knapen:2017xzo}.} An estimation of $\sigma^{\text{max}}_{\chi e}$ for scattering off electrons and a corresponding HYPER model coupled to electrons would be relevant for a host of proposed electron-recoil-based future experiments. An electron HYPER would have a dark phase transition temperature below $m_{e}$. It is an interesting question as to whether such a low phase transition temperature could modify or remove bounds on the mediator from HB stars. We leave a detailed study of the possible models and associated signals and constraints to future work ~\cite{future}. 

\begin{acknowledgments}
We thank Prudhvi N.~Bhattiprolu, Tim Cohen, Fatemeh Elahi, Joshua Foster, Simon Knapen, Gordan Krnjaic, Robert Lasenby, Filippo Sala, Katelin Schutz, Pedro Schwaller, Juri Smirnov, and Yuhsin Tsai for useful discussions. The research of GE is supported by the Cluster of Excellence {\em Precision Physics, Fundamental Interactions and Structure of Matter\/} (PRISMA${}^+$ -- EXC~2118/1) within the German Excellence Strategy (project ID 39083149). RM and AP are supported in part by the DoE grant DE-SC0007859. AP would like to thank the Simons Foundation for support during his sabbatical. RM thanks the Galileo Galilei Institute for Theoretical Physics (GGI) and the Mainz Institute for Theoretical Physics (MITP) of the Cluster of Excellence PRISMA${}^+$ (project ID 39083149) for their hospitality while a portion of this work was completed.

\end{acknowledgments}
\bibliography{Refs}

\clearpage

\onecolumngrid
\begin{center}
  \textbf{\large Supplementary Material for Maximizing Direct Detection with HYPER Dark Matter}\\[.2cm]
  \vspace{0.05in}
  {Gilly Elor, \ Robert McGehee, \ and \ Aaron Pierce}
\end{center}

\twocolumngrid
\setcounter{equation}{0}
\setcounter{figure}{0}
\setcounter{page}{1}
\makeatletter
\renewcommand{\theequation}{S\arabic{equation}}
\renewcommand{\thefigure}{S\arabic{figure}}

In this supplementary material, we calculate the $\gamma \gamma \to \phi$ rate and derive a sufficient condition to prevent these inverse decays from increasing the $\chi$ relic abundance after the phase transition. We then discuss the considerations needed to explore the HYPER parameter space for $\TPT > 1 \, \text{MeV}$ and show the HYPER space for a $\TPT = 5 \text{ MeV}$ benchmark. We also include the cross sections for $\bar{\chi} \chi \to \phi \phi$ and $\bar{\chi} \chi \to \gamma \gamma$ for reference. 

\section{\texorpdfstring{$\gamma \gamma \to \phi$}{gam gam to phi} Inverse Decays}
\label{sec:gamgam2phi}

We address the $\gamma \gamma \to \phi$ inverse decays here in more detail. Before the phase transition, $m_\phi^i \gg T$ for our UV freeze-in scenario. But after the phase transition, $m_\phi^\text{min} < \TPT$, so $\gamma \gamma \to \phi$ inverse decays may occur. The decay rate in the $\phi$ rest frame is
\al{
\label{eq:phi2gamma}
\Gamma_{\phi \to \gamma \gamma} = \prn{\frac{\alpha y_n}{4 \pi m_n}}^2 \frac{m_\phi^3}{2\pi}.
}

For the moment, we assume the $\gamma \gamma \to \phi$ inverse decays are the dominant, $\phi$-number changing process. The $\phi$ number density Boltzmann equation then is
\al{
\label{eq:phinumdens}
\frac{\partial n_\phi}{\partial t} + 3Hn_\phi = \avg{\Gamma_{\phi \to \gamma \gamma}} n_\phi^\text{eq},
}
where the thermal average and equilibrium $\phi$ number density are both evaluated at the SM bath temperature. 

These inverse decays must be prevented for different reasons depending on $m_\chi$. Let us first consider the scenario in which $m_\phi < 2 m_\chi$. In this case, any $\phi$s which are produced by the SM bath after the phase transition will not go on to produce DM via decays. Instead, they will just eventually decay back to SM photons (or to lighter dark-sector degrees of freedom, if any are present and coupled to $\phi$). In this case, we need only make sure that $\phi$s never come close to thermal equilibrium to prevent them from contributing noticeably to $\Neff$ at BBN. Thus, the condition we impose from Eq.~\eqref{eq:phinumdens} is simply
\al{
\avg{\Gamma_{\phi \to \gamma \gamma}} \ll 3H.
}
We find that this is easily satisfied for $m_\phi = m_\phi^\text{min}$ and $y_n = y_n^\text{max}$.

If, however, $m_\phi \ge 2 m_\chi $, then any produced $\phi$s can quickly decay to DM pairs and increase the DM relic abundance after the phase transition. Approximating the equilibrium $\phi$ distribution as Maxwell-Boltzmann, we thus require 
\al{
\label{eq:gamgam2phicond}
\Gamma_{\phi \to \gamma \gamma} \frac{m_\phi^2 T}{2\pi^2} K_1 (m_\phi / T) \lesssim 0.15 H n_{\chi}\,.
}
This ensures subsequent $\phi$ decays to DM pairs do not increase the DM number density by more than $10\%$. Setting $y_n=y_n^\text{max}$ in Eq.~\eqref{eq:phi2gamma}, we find that we need $m_\phi \gtrsim 21 \text{ MeV}$ in order to satisfy Eq.~\eqref{eq:gamgam2phicond} for $\TPT = 1 \text{ MeV}$. If we consider instead that $\TPT = 5 \text{ MeV}$, as we do in the next section, the constraint yields $m_\phi \gtrsim 110 \text{ MeV}$. Thankfully, these inverse decays only become an issue for lighter DM masses. 

\section{HYPERs at Higher $\TPT$}

In the body of this paper and Fig.~\ref{fig:sigmax}, we have chosen the benchmark value $\TPT = 1 \text{ MeV}$. But it is interesting to consider what $\sigma_{\chi n}$ HYPERs may be able to achieve if the phase transition occurred at higher temperatures. In this case, there are a number of additional processes that could impact the HYPER abundance.

\begin{figure}[t!]
\centering
\includegraphics[width=0.5 \textwidth]{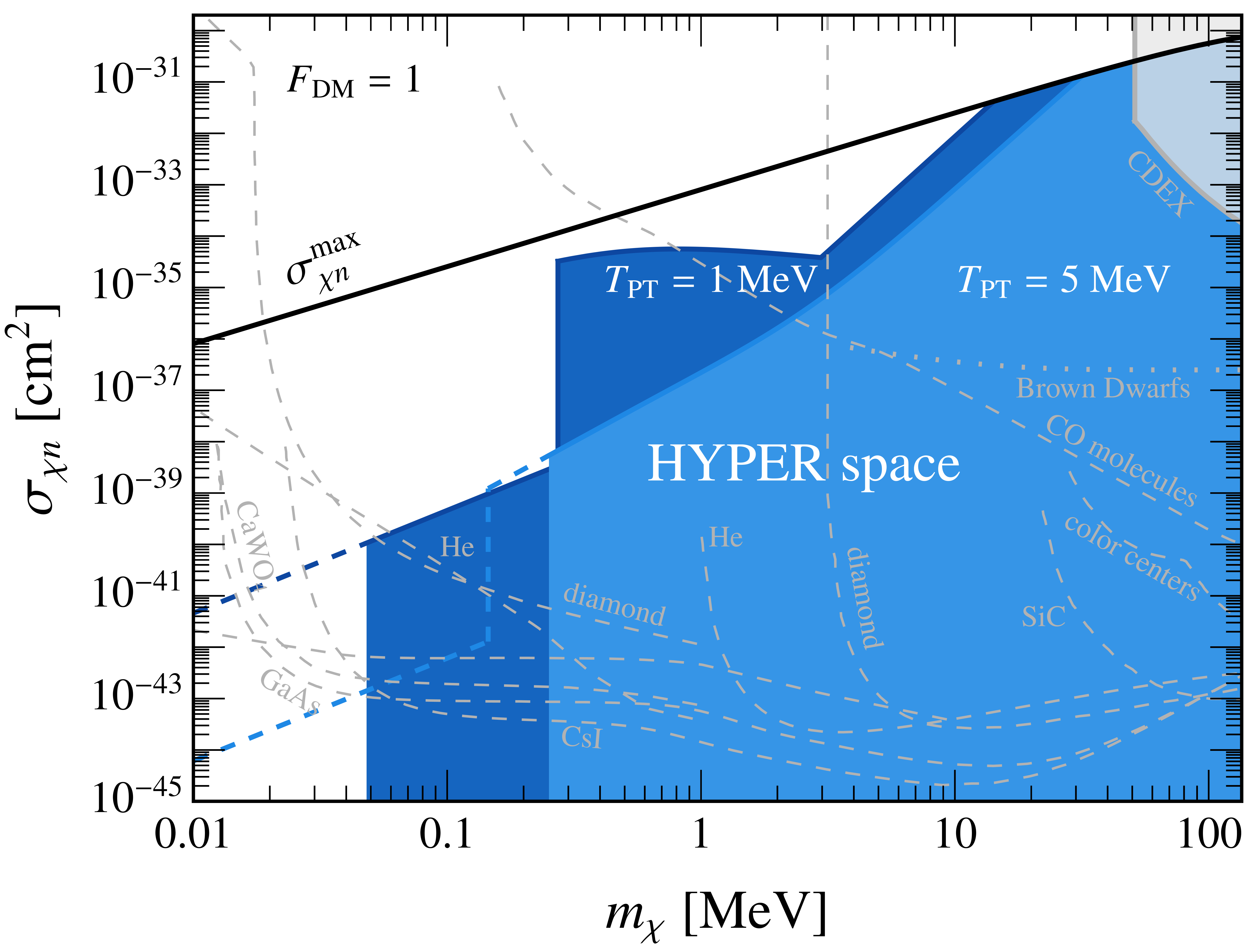}
\caption{The HYPER parameter space for the $\TPT=1\, \text{MeV}$ benchmark (as shown in the  body of the paper), along with the $\TPT=5\, \text{MeV}$ benchmark which represents the largest phase transition temperature for which pion interactions are not important. The current constraint from CDEX~\cite{CDEX:2019hzn} is shaded gray, while future projected sensitivities are shown with dashed gray lines \cite{Knapen:2016cue,Budnik:2017sbu,Kurinsky:2019pgb,Essig:2019kfe,Trickle:2019nya,Griffin:2019mvc,Campbell-Deem:2019hdx,Griffin:2020lgd,Coskuner:2021qxo,Leane:2020wob}.}
\label{fig:HYPERs_TPT5MeV}
\end{figure}

For larger values of $\TPT$, even though their abundance in the bath can be substantially suppressed, pions may still be present in  numbers sufficient to enable production of $\phi$s or $\chi$s after the phase transition.  These would disturb the DM abundance and/or lead to additional constraints from $\Neff$. $\phi$ production processes involving $\pi$s include $\pi \pi \to \phi \phi$, $\pi^+ \pi^- \to \phi \gamma$, $\pi^\pm \gamma \to \pi^\pm \phi$, and $\pi^\pm \gamma \to \pi^\pm \phi \gamma$. And for each of the above, a $\phi$ could be replaced by a $\bar{\chi} \chi$ pair to give a DM production process that may be important. These processes are mediated by the couplings (see \cite{Gunion:1989we}, and references therein) 
\bea
\label{eq:Lphipipi}
\mathcal{L} \,\supset\,  \frac{3 y_n }{m_n}  \left( \frac{2}{3}  \phi  \left|D^\mu \pi^+\right|^2 - m_\pi^2 \phi \pi^+ \pi^-  \right) \,.
\eea

While it is beyond the scope of this work to do a detailed analysis of the above processes, we can get a feeling for the values of $T_{PT}$ at which they become relevant. Out of all of the above processes, $\pi^\pm \gamma \to \pi^\pm \phi$ produces a non-negligible amount of $\phi$s at the lowest $\TPT$.  So, let us estimate the value of $\TPT$ at which this process starts producing an appreciable number of  $\phi$s.

For $T << m_{\pi}$, the approximate cross section for $\pi^\pm \gamma \to \pi^\pm \phi$ is
\al{
\sigma v_{\pi^+ \gamma \to \pi^+ \phi} \approx \frac{23}{12} \frac{\alpha y_n^2}{m_n^2}.
}
We require that
\al{
\sigma v_{\pi^+ \gamma \to \pi^+ \phi} \, n_\gamma^{\text{eq}} n_{\pi^+} ^{\text{eq}} \lesssim 0.15 H n_\chi.
}
As with Eq.~\eqref{eq:gamgam2phicond} above, this ensures subsequent $\phi$ decays to DM pairs do not increase the DM number density by more than $10\%$. We find that this condition is satisfied for all $\TPT \lesssim 5.2 \text{ MeV}$. 

So, for illustrative purposes, we calculate the maximum $\sigma_{\chi n}$ for HYPERs assuming a phase transition at $\TPT = 5 \text{ MeV}$ while only imposing the HYPER constraints discussed in the body of the paper. The result is shown as the light blue region in Fig.~\ref{fig:HYPERs_TPT5MeV} overlaid on-top of the $\TPT = 1 \text{ MeV}$ from Fig.~\ref{fig:sigmax}. In addition to this new $\TPT$ contour, we have extended the $\TPT = 1 \text{ MeV}$ contour from $m_\chi = \omega_p(\TPT)/2=48 \text{ keV}$ down to $m_\chi = 10 \text{ keV}$ with a dashed line. For these lighter DM masses, it may be possible for longitudinal plasmon modes mixing with $\phi$ to decay to $\bar{\chi} \chi$ pairs. Thus, for these lighter $m_\chi$, it may be necessary to impose additional constraints on $\prn{y_n,y_\chi,m_\phi}$ to prevent too much DM production after the phase transition due to plasmons. To our knowledge, plasmon decays to DM pairs in the presence of a scalar mediator have not been considered in the literature and is beyond the scope of this work. 

Since we already described the prescription for choosing $(y_\chi,m_\phi)$ for $\TPT = 1 \text{ MeV}$, we describe here the analogous choices for the $\TPT = 5 \text{ MeV}$ contour which result in the largest HYPER $\sigma_{\chi n}$, shown as the boundary of the light blue region in Fig.~\ref{fig:HYPERs_TPT5MeV}.

\begin{itemize}
\item $5 \text{ MeV} < m_\chi < m_{\pi^0}$: $\bar{\chi}\chi \rightarrow \phi \phi$ is the most constraining process; maximized $\sigma_{\chi n}$ results from setting $m_\phi = m_\phi^\text{min}$ and taking $y_\chi < y_\chi^\text{max}$ to satisfy Eq.~\eqref{eq:chichi2phiphiCond}.\\

\item $m_\phi^\text{min}/2 \text{ MeV} < m_\chi < 5 \text{ MeV} $: both $\bar{\chi} \chi \to \phi \phi$ and $\gamma \gamma \to \bar{\chi}\chi$ are constraining, since $m_\chi \le \TPT$; maximized $\sigma_{\chi n}$ results from setting $m_\phi = m_\phi^\text{min}$ and $y_\chi < y_\chi^\text{max}$ such that both Eqs.~(\ref{eq:gamgam2chichiCond}-\ref{eq:chichi2phiphiCond}) are satisfied; we dash the HYPER curve for $m_\chi \le \omega_p(\TPT)/2= 250 \text{ keV}$ since it is possible that longitudinal plasmons mixing with $\phi$ may allow for $\gamma^\ast \to \bar{\chi} \chi$ decays \cite{Dvorkin:2019zdi}, which we have not calculated.

\item $m_\chi < m_\phi^\text{min}/2 = 150 \text{ keV}$: $\phi \to \bar{\chi} \chi$ is always possible, so $\gamma \gamma \to \phi$ must be kinematically forbidden by setting $m_\phi = 110 \text{ MeV}$, as discussed above; maximized $\sigma_{\chi n}$ results from then setting $y_\chi = y_\chi^\text{max}$; we have checked that Eq.~\eqref{eq:gamgam2chichiCond} is satisfied.
\end{itemize}

\section{Cross Sections for Dark Matter annihilations \label{sec:chichitophiphi}} 
The spin-averaged matrix element squared, which agrees with the result in \cite{Scott:2016abc}, for $\bar{\chi} \chi \to \phi \phi$ is
\al{
\frac{1}{4} \sum_\text{spins} \left|\mathcal{M}\right|^2 = 32 y_\chi^4 \prn{E_\chi^2-m_\chi^2} \frac{a}{b^2}\,,
}
where
\al{
a = &\ctheta{2} \prn{E_\chi^2-m_\phi^2}\prn{E_\chi^4+2m_\chi^2 m_\phi^2 - E_\chi^2 \prn{4m_\chi^2+m_\phi^2}} \nonumber \\
&-\ctheta{4} \prn{E_\chi^2-m_\chi^2} \prn{E_\chi -m_\phi}^2 \prn{E_\chi +m_\phi}^2 \\
&+m_\chi^2 \prn{m_\phi^2-2E_\chi^2}^2 , \nonumber \\
b= &\,4(1-\ctheta{2})E_\chi^4-4\ctheta{2} m_\chi^2 m_\phi^2 + m_\phi^4 \\
&+4 E_\chi^2 \prn{-m_\phi^2 + \ctheta{2} \prn{m_\chi^2+m_\phi^2}} , \nonumber
}
and its cross section is
\al{
\sigma v_{\bar{\chi} \chi \to \phi \phi} = \frac{\sqrt{E_\chi^2-m_\phi^2}}{64\pi E_\chi^3} \int_{-1}^{1} d (\cos \theta) \frac{1}{4} \sum_\text{spins} \left|\mathcal{M}\right|^2.
}

The cross section for $\chi \bar{\chi} \rightarrow \gamma \gamma$, relevant for indirect detection, is
\al{
\sigma v_{\bar{\chi} \chi \to \gamma \gamma} \! = \! \frac{1}{8 \pi} \! \prn{\frac{ \alpha y_n y_\chi}{2 \pi m_n}}^2 \! \frac{s \prn{s-4m_\chi^2}}{\prn{s-m_\phi^2}^2}.
\label{eq:chichi2gamgam}
}

\end{document}